\documentclass[12pt]{iopart}
\usepackage{iopams}
\usepackage{setstack}
\usepackage{braket}

\usepackage{graphicx}
\usepackage{dcolumn}
\usepackage[utf8]{inputenc}
\usepackage[english]{babel}
\usepackage{float}
\usepackage{subfigure}
\usepackage{color}
\usepackage{pgfplots} 
\usepackage{tikz}
\usepackage{multirow}
\usepackage{booktabs} 
\usepackage{colortbl}
\usepackage{array}

\newcommand{\eqref}[1]{(\ref{#1})}
\newcommand{\figref}[1]{\figurename~\ref{#1}}
\newcommand{\secref}[1]{Section \ref{#1}}
\makeatletter
\def\l@subsubsection#1#2{}

\makeatother

\pgfplotsset{compat=1.8}
\begin{document}


\title[Numerical study of the free-energy
  barriers of spin glasses]{Numerical study of barriers and valleys in the free-energy
  landscape of spin glasses}

\author{I.~González-Adalid Pemartín$^1$, V.~Martin-Mayor$^{1, 2}$, G.~Parisi$^3$ and J.J.~Ruiz-Lorenzo$^{4, 5, 2}$}

\address{$^1$ Departamento  de F\'\i{}sica Te\'orica, Universidad Complutense, 28040 Madrid, Spain}
\address{$^2$ Instituto de Biocomputaci\'on y F\'{\i}sica de Sistemas Complejos (BIFI), 50018 Zaragoza, Spain}
\address{$^3$ Dipartimento di Fisica, Sapienza
  Universit\`a di Roma, INFN, Sezione di Roma 1, and CNR-Nanotec,
  I-00185 Rome, Italy}

\address{$^4$ Departamento de F\'{\i}sica, Universidad de Extremadura, 06006 Badajoz, Spain}
\address{$^5$ Instituto de Computaci\'on Cient\'{\i}fica Avanzada (ICCAEx), Universidad de Extremadura, 06006 Badajoz, Spain}


\begin{abstract}
We study the problem of glassy relaxations in the presence of an
external field in the highly controlled context of a spin-glass
simulation.  We consider a small spin glass in three dimensions
(specifically, a lattice of size $L=8$, small enough to be
equilibrated through a Parallel Tempering simulations at low
temperatures, deep in the spin glass phase). After equilibrating the
sample, an external field is switched on, and the subsequent dynamics
is studied. The field turns out to reduce the relaxation time, but
\emph{huge} statistical fluctuations are found when different samples
are compared. After taking care of these fluctuations we find that
the expected linear regime is very narrow. Nevertheless, when regarded
as a purely numerical method, we find that the external field is extremely
effective in reducing the relaxation times. 
\end{abstract}


\submitto{\jpa}                              
                              
\maketitle


\section{\label{sec:Intro}Introduction}

The glass transition poses a challenge for modern physics, both
theoretically and experimentally
\cite{berthier:11,cavagna:09}. Indeed, upon cooling, the relaxation
times of a large variety of systems (supercooled liquids, colloids,
type-II superconductors, etc.) grow beyond bound. This dynamic arrest
could seem characteristic of a phase transition, were it not for the
absence of any accompanying change in structural properties. 

This state of affairs poses a paradox. Imagine that (in analogy with
the Swendsen-Wang method for ferromagnets~\cite{swendsen:87}) a
beautiful Monte Carlo sampling method is invented, bearing no
relationship whatsoever with natural dynamics but producing
\emph{equilibrium} configurations at a very low temperature. We are
referring to such a low temperature that no Nature-imitating dynamics
could equilibrate the system in any reasonable simulation time (for
instance, equilibration would be impossible with a Molecular Dynamics
simulation). In fact, to some extent, swap methods for simple
glass-formers~\cite{grigera:01,fernandez:06,ninarello:17} or Parallel
Tempering dynamics for spin-glasses~\cite{hukushima:96,marinari:98b}
put us in such situation. Now, how could one use these otherwise
unreachable equilibrium configurations?

As we said above, structurally the system does not change
significantly with temperature, hence computing static observables
(such as particle densities or structure factors) would be somewhat
uninteresting. Clearly, one needs to consider some Nature-imitating
dynamics, but using our unphysically-obtained equilibrium
configurations as initial conditions. However, not only the natural
dynamics at low-temperature is slow, it also
\emph{ages}~\cite{struik:80,vincent:97}. When a glass is left to
evolve at low temperatures, the longer the relaxation the slower the
dynamic responses (a.c. dielectric or magnetic susceptibilities, for
instance), and our equilibrated starting configurations are
effectively equivalent to a natural system that has been left to relax
for an exceedingly long waiting-time.

We propose here a general solution for the above outlined problem and
demonstrate its feasibility in the particular case of
spin-glasses~\cite{mezard:87,young:98}.

\begin{figure}[htbp]
\subfigure[\hspace{0.5mm} Without
  field]{\includegraphics[width=0.49\linewidth]{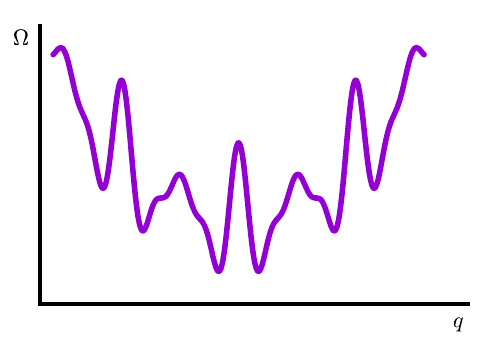}\label{fig:E_sin_campo}}
\subfigure[\hspace{0.5mm} With
  field]{\includegraphics[width=0.49\linewidth]{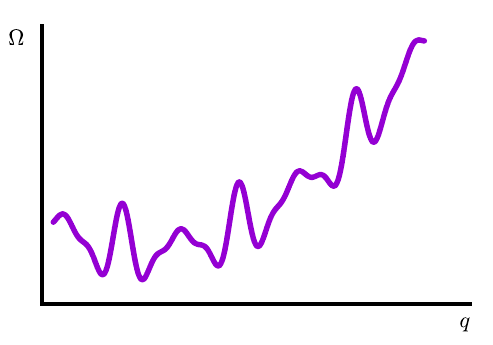}\label{fig:E_con_campo}}
\caption{Cartoon of the free-energy landscape without (a) and with (b)
  external field. We name $\Omega$ to the free energy, and use the
  correlation $q$ (see \secref{sub:Obs}) as collective coordinate, in
  order to make the simplest possible representation.}
\label{fig:E_con_sin_campo}
\end{figure}

Our starting point is the physical view of a rugged landscape with
many local minima of the free-energy, separated by
barriers~\cite{berthier:11,cavagna:09} Now, let $q$ be an appropriate
collective coordinate describing the landscape [see \secref{sub:Obs}
  and \figref{fig:E_sin_campo}]. Upon coupling $q$ to an external
field of strength $\varepsilon$, we expect the time needed to scape
from a local minimum to vary as
\begin{equation}
\tau^{(\varepsilon)} \propto \exp\left[-\frac{N\Delta q \varepsilon}{T}\right] \ ,
\label{eq:tau_eps}
\end{equation}
where $N$ is the number of particles in the system, and $\Delta q$ is
the typical width of the barrier (we shall provide a more precise definition for $\tau^{(\varepsilon)}$ below). Clearly, the benefits of
equation~\eqref{eq:tau_eps} are twofold. On the one hand, the time needed
to relax the system may be enormously smaller than its $\varepsilon\to
0$ limit (thus making the simulation feasible). And, on the other
hand, information on the barriers can be gained by considering the
$\varepsilon$-dependence.

Let us mention that this strategy has been followed experimentally, by
applying an uniform magnetic field to a spin-glass sample, in order to
extract the spin-glass correlation length~\cite{joh:99,guchhait:17}
(the $\varepsilon$ considered here is proportional to the
\emph{square} of the external magnetic field~\cite{janus:17b}). Very
similar approaches are currently under investigation for the
dielectric response of glass-forming liquids~\cite{lhote:14,ladieu:12}.

We have chosen spin-glasses to study the effect of adding a field, for
a variety of reasons: (i) spin glasses are unique among glass formers
in that we know that their glassy dynamics arises from a~\emph{bona
  fide} phase
transition~\cite{gunnarsson:91,palassini:99,ballesteros:00}; (ii)
numerical simulations of spin-glasses are so simple that
special-purpose hardware has been
built~\cite{ogielski:85,janus:08,janus:14}; (iii) high experimental accuracy
(thanks to the SQUID) has given experimental access to
non-linear susceptibilities long time ago~\cite{gunnarsson:91} (the
equivalent studies in supercooled liquids such as glycerol are only
incipient nowadays~\cite{albert:16}) and (iv) we understand better
than in any other glassy model-system how macroscopic responses relate
with microscopic correlation functions that we can compute in the
simulations~\cite{janus:17b,janus:18}.

The layout of the remaining part of this paper is organized as follows.
In \secref{sub:Model} we introduce the model we study and the
corresponding observables, before explaining our methods in
\secref{sec:Met}. In particular, we explain the protocol in
\secref{sub:protocolo}, while the quantities used to compute $\tau$ in
\secref{sub:Obs}. In \secref{sec:ResulyDisc}
we show the results and the relation between $\tau$ and the field
$\varepsilon$ \secref{sub:Resul}. We explain some effects that virtual
states might produce in \secref{sub:Virtual}. Finally, our conclusions
are given in \secref{sec:Concl}.  \ref{Ap:Sim} contains 
technical details.

\section{\label{sub:Model}Model and observables}
We consider the Edwards-Anderson model~\cite{edwards:75,edwards:76} on a cubic lattice with periodic boundary conditions. The Hamiltonian is
\begin{equation}
\mathcal{H}=-\sum_{\langle i,j\rangle}J_{ij}\sigma_i\sigma_j\ ,
\label{eq:H_LI}
\end{equation} 
where $\langle i,j\rangle$ is a sum over lattice nearest-neighbours,
$J_{ij}$ is the coupling constant for the bond joining lattice-nodes
$i$ and $j$, and $\sigma_i=\pm 1$ is the spin at the $i$-th node of
the lattice. We set independently for each lattice bond 
$J_{ij}=\pm 1$ with 50\% probability. Let us remark that the
$\{J_{ij}\}$ are static variables (quenched disorder), while the
$\{\sigma_i\}$ are dynamic. Hence, one should perform \emph{first} the
dynamic averaging for a fixed set of couplings (denoted $\langle
\ldots\rangle$ hereafter). The averaging over the coupling values is
only performed afterwards.

It is well known that model \eqref{eq:H_LI} is endowed
with a $Z_2$ gauge-invariance~\cite{toulouse:77}: $\mathcal{H}$
remains unchanged after the transformation $\sigma_i\rightarrow
\eta_i\sigma_i,\ J_{ij}\rightarrow \eta_iJ_{ij}\eta_j$ (where
$\eta_i=\pm 1$ is set independently for each lattice
site). Furthermore, the probability for the original couplings
$\{J_{ij}\}$ is identical to the probability of their gauge-transform
$\{ \eta_iJ_{ij}\eta_j\}$. Therefore, meaningful observables should be
gauge-invariant as well. In particular, the building block for our
observables will be the two-times \emph{overlap} that compares the
value of the very same spin at times $t_1$ and $t_2$:
\begin{equation}
q_{i}(t_1,t_2)=\sigma_i(t_1)\sigma_i(t_2)\,.
\label{eq:overlap}
\end{equation}
[to lighten the notation, we shall omit the argument $(t_1,t_2)$
whenever it will be possible; see \secref{sub:Obs} for a discussion
of the limit $t_1-t_2 \rightarrow \infty$].

In addition, we have introduced a external field which tries to drive
away the system from an already-equilibrated initial configuration,
thus accelerating the relaxation process. For this, we added to the
Hamiltonian \eqref{eq:H_LI} a repulsion term
\begin{equation}
\mathcal{H}=-\sum_{\langle i,j\rangle}J_{ij}\sigma_i\sigma_j\ +\ 
\varepsilon\sum_i\sigma_i(t=0)\sigma_i(t)\,,
\label{eq:repulsion}
\end{equation}
where $\varepsilon$ term can be regarded as a locally varying external
magnetic
field $h_i\equiv\varepsilon \sigma_i(t=0)$. Of course, the larger
$\varepsilon$, the stronger becomes the repulsion from the (equilibrium)
starting condition.

\section{\label{sec:Met}Methods}

Let us recapitulate two problems that we need to address. 

First, as it is clear from~\eqref{eq:repulsion}, we need to reach
thermal equilibrium at a low temperature, deep in the glassy phase. We
shall solve this problem using a Parallel Tempering
algorithm. Unfortunately, the computational resources needed to
equilibrate the system grow very fast with the lattice size
$L$~\cite{janus:10,billoire:18}. This difficulty, together with the
need of studying a large number of samples, has convinced us to
restrict ourselves to $L=8$ in this exploratory study.

Second, we need to simulate the $\varepsilon$-dependent Metropolis
dynamics at fixed temperatures from those starting
configurations. Because this Nature-imitating dynamics should be
followed for very long times (for small values of $\varepsilon$, at
least), we have found it convenient to use properly adapted multispin
coding techniques.

Section~\ref{sub:protocolo} explains our solutions to both problems
(more details can be found in \ref{Ap:Sim}). The observables that we
considered are addressed in Section~\ref{sub:Obs}.

\subsection{\label{sub:protocolo}Simulations Protocol}

We have divided the simulation process in three parts:

\paragraph{Thermalization process.} We need to equilibrate a large number of samples at a low temperature. We have chosen as working temperature
$T=0.698\approx 0.63 T_\mathrm{c}$ (recall that
$T_\mathrm{c}=1.1019(29)$ is the critical temperature separating the
paramagnetic from the spin-glass phase~\cite{janus:13}).  We have
reached equilibrium at $T=0.698$, through a Parallel Tempering
simulation, for 1280 samples.

Specifically, our Parallel Tempering simulation for each sample
contains 13 clones evenly distributed in the temperature interval
$\left[0.698,1.575\right]$. For every clone, we performed 10
Metropolis full-lattice sweeps at fixed temperature. After that, we
performed a temperature-swap attempt. We took a total of $2.5\times
10^8$ Metropolis sweeps per clone (this simulation time is longer by a
factor of $5\times 10^4$ than the longest temperature-mixing time
identified in Ref.~\cite{billoire:18} for $L=8$). Once the Parallel
Tempering simulation was completed, we took as initial configuration
for the next step in our protocol the temperature-clone that occupied
the lowest temperature (namely $T=0.698\approx 0.63 T_\mathrm{c}$).

In practice, we grouped our 1280 samples in 10 bunches of 128 samples
each.  The 128 samples in a bunch were simulated in parallel using a
standard multi-sample multispin coding~\cite{newman:99}. Indeed, we
coded on a single 128-bits computer word the spins that occupied the
very same lattice site for each of the 128 samples in a bunch. As it
is usual in multi-sample multispin coding, during the Metropolis part
of the simulation we used only one random number per computer word. On
the other hand, for the temperature-swap attempt we employed, of
course, 128 independent random-numbers for the 128 samples in a bunch.

\paragraph{Overcoming possible correlations.} 
After the Parallel Tempering, we can assume that the temperature-clone
at the lowest temperature for each of our $1280$ samples has been randomly
extracted from our target distribution:
\begin{equation}
  P(\left\lbrace\sigma_i\right\rbrace)= P(\left\lbrace
  J_{ij}\right\rbrace) P_B(\left\lbrace\sigma_i\right\rbrace |\left\lbrace
  J_{ij}\right\rbrace) \,,
  \label{eq:dis_prob_t=0}
\end{equation}
where $P(\left\lbrace J_{ij}\right\rbrace)$ is the uniform
distribution for the coupling constants, while
$P_B(\left\lbrace\sigma_i\right\rbrace |\left\lbrace
J_{ij}\right\rbrace)$ is the conditional Boltzmann distribution of the
spin configuration given the couplings.

However, we note that the Metropolis part of the Parallel Tempering simulation
might induce some correlation among the 128 samples in a bunch, due to
the sharing of random numbers. In order to alleviate this problem, we
have further simulated the temperature clone at the lowest
temperature, $T=0.698$, for some additional $2\times 10^{10}$
Metropolis steps. At such a low temperature, we can use an independent
random-number for every bit in our computer word with almost no
computational overhead~\cite{fernandez:15} (see
also~\ref{Ap:Sim}). 

After this de-correlating step, we finally have a starting
configuration $\{\sigma_i(t=0)\}$ for each of our 1280 (statistically independent) samples.

\paragraph{Dynamics with and without field.} 
Finally, for every value of $\varepsilon$, recall
Equation~\eqref{eq:repulsion}, and from every one of our 1280
starting configurations, we simulated $49$ independent trajectories (or
replicas). The rationale for such a large number of trajectories per
starting point is explained in \secref{sub:Obs}.  We performed
Metropolis dynamics at $T=0.698$. Details on our choice of Metropolis
dynamics and our multispin coding algorithm can be found
in~\ref{Ap:Sim}. The parameters characterizing these simulations are
summarized in Table \ref{tabla:para_sim}.

\begin{table}[t]
\centering
\begin{tabular*}{\columnwidth}{@{\extracolsep{\fill}}|c|ccccccc|}
\hline
      $\varepsilon$ & 0 & 0.0005 & 0.001 & 0.003 & 0.006 & 0.01 & 0.02 \\
\hline
      Metropolis sweeps ($\times10^8$) & 200 & 140 & 100 & 26 & 3.2 & 0.2 & 0.16\\ \hline
\end{tabular*}
  \caption{For each of our 1280 samples, and for every value of
    $\varepsilon$ [recall Equation~\eqref{eq:repulsion}], we give the
    total number of Metropolis sweeps that we took along every one of
    our 49 trajectories.  All 49 trajectories started from the same,
    already equilibrated, starting configuration $\{\sigma_i(t=0)\}$
    (see Section~\ref{sub:protocolo}). We sampled the state
    $\{\sigma_i(t)\}$, and computed the corresponding observables,
    400000 times for each trajectory. The sampling times were evenly
    distributed along the duration of the run.}
\label{tabla:para_sim}
\end{table}

\subsection{\label{sub:Obs}Observables}
Our main goal is computing the time scale $\tau^{(\varepsilon)}$,
recall Equation~\eqref{eq:tau_eps}, in a meaningful way. As we will
show below by example, we will be hampered by severe statistical
fluctuations (of several types). We shall explain these difficulties
and our solutions to overcome them.

\begin{figure}[htbp]
\centering
\includegraphics[width=0.9\linewidth]{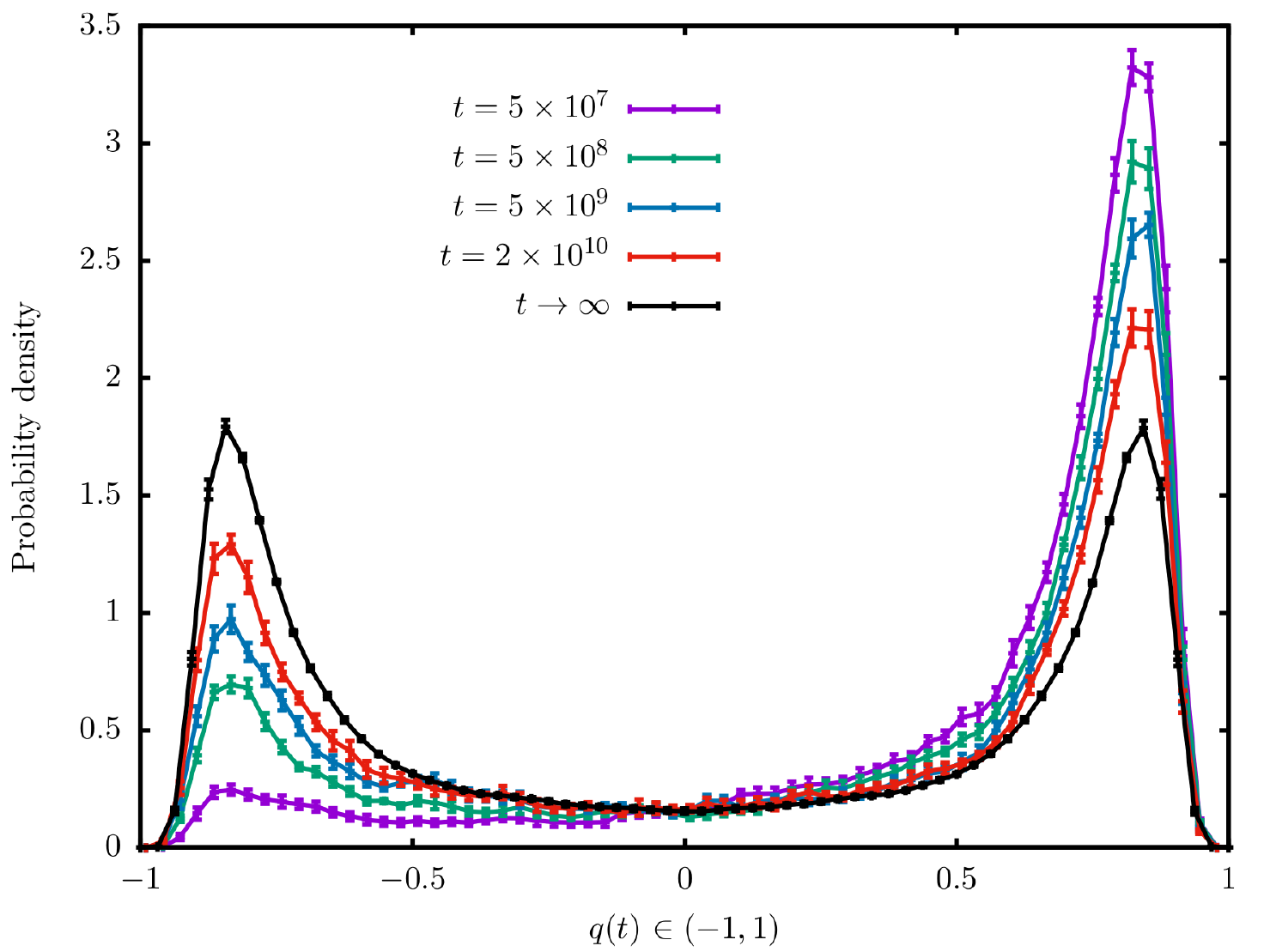}
\caption{Probability density of the correlation $q(t)$, see
  Equation~\eqref{eq:correlacion}, as computed at different times for
  $\varepsilon=0$ over our 1280 samples and starting points. In the
  limit $t\to\infty$, the configurations $\{\sigma_i(t)\}$ and
  $\{\sigma_i(t=0)\}$ becomes statistically independent and the
  distribution becomes symmetric $P[q(t\to\infty)]=P[-q(t\to\infty)]$,
  thus reflecting the symmetry of the Hamiltonian in
  Equation~\eqref{eq:repulsion} when $\varepsilon=0$. Indeed, as the
  time proceeds, the asymmetry in the probability density $P[q(t)]$
  decreases.  The limit $t\to\infty$ was obtained through Parallel
  Tempering (we equilibrated 8 independent replicas and computed the
  resulting 28 overlaps, for each of our 1280 samples).}
\label{fig:DistribProb}
\end{figure}

\paragraph{Our basic quantity: the time correlation (also known as overlap).}
Given the sluggish dynamics in a glassy phase, it is likely that if we
observe an spin now, and we look at it again some time later, there is
a sizeable probability that the spin remains in the same state
\cite{edwards:75}. This effect is quantified by the time correlation
function, recall Equation~\eqref{eq:overlap}, with the (already in
equilibrium) initial configuration\footnote{The reader should not
  confuse our $q(t)$ with the analogous quantity computed in
  equilibrium simulations, where the overlap is computed between
  statistically independent configurations. Indeed, our $q(t)$ becomes
  the standard overlap only in the limit $t\to\infty$,
  see~\figref{fig:DistribProb}.}
\begin{equation}
q(t)=\frac{1}{N}\sum_i\sigma_i(t=0)\sigma_i(t)\,,
\label{eq:correlacion}
\end{equation}
where $N=8^3$ is the total number of spins in our system.
Note that the maximum value for the overlap is $q(t)=1$ (i.e. no spin has changed), while the minimum value is $q(t)=-1$ (i.e. every spin has changed).

The probability distribution of $q(t)$, $P[q(t)]$, as computed over our 1280
samples for $\varepsilon=0$, unveils extremely slow relaxations and
exposes large statistical fluctuations, see
Figure~\ref{fig:DistribProb}.

As for the slow dynamics, let at recall that at infinite time
$P[q(t)]$ is symmetric $P[q]=P[-q]$ due to the global spin-flip
symmetry at $\varepsilon=0$. Nevertheless, see
\figref{fig:DistribProb}, for small $t$ the distribution is strongly
peaked at the Edwards-Anderson parameter for $T=0.698$ and system size
$L=8$, namely $q_{\mathrm{EA}}\approx 0.844$. As time goes by, $P[q(t)]$
develops a secondary peak at $q(t)=-q_{\mathrm{EA}}$. This secondary
peak grows with time. However, even after $2\times10^{10}$ Metropolis
sweeps, we still have $P[q(t)=-q_{\mathrm{EA}}]\approx 0.6
P[q(t)=+q_{\mathrm{EA}}]$. This result implies~\cite{sokal:97} that
the time necessary to reach thermal equilibrium with a Metropolis
dynamics at $T=0.698$ is enormously longer than $2\times 10^{10}$
full-lattice sweeps. As we anticipated in the Introduction,
equilibrating this system at such a low temperature is feasible
nowadays only with a non-physical Monte Carlo dynamics, such as
Parallel Tempering.

Besides, we observe in \figref{fig:DistribProb} that even at very
short times there is a sizeable probability of finding \emph{any}
value of $q(t)$.  These large fluctuations make unpractical the
straightforward definition of the relaxation time $\tau$
\begin{equation}
q(t=\tau)=0\,.
\label{eq:cond_q(tau)}
\end{equation}
In order to overcome this problem, we shall need to consider statistical
fluctuations in greater details.

\begin{figure}
\centering
\includegraphics[width=0.9\linewidth]{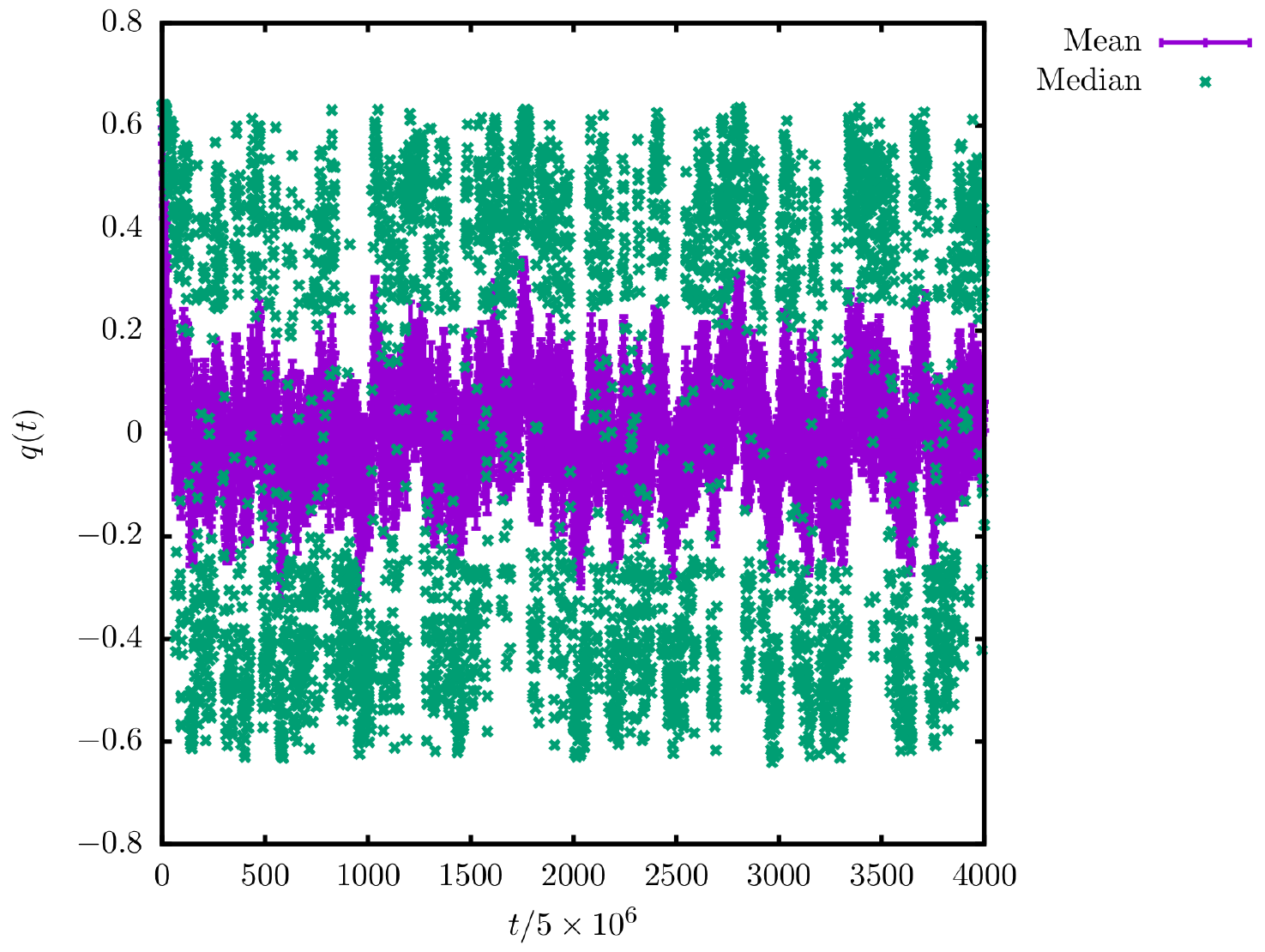}
\caption{Time evolution of the median and the mean of the conditional
  probability distribution function
  $P[q(t)\,|\,\{J_{ij},\sigma_i(t=0)\}]$, defined in
  Equation~\eqref{eq:PJsigma0}, as computed from the 49 replicas for
  our sample \#107 at $\varepsilon=0$. We show errorbars for the mean. Errorbars for
  the median are as large as the median itself (indeed, the median is
  ill-defined for probability density functions with two bands, each
  of them with the same weight).}
\label{fig:fluctuaciones_q}
\end{figure}
\paragraph{Fluctuations within the same sample and starting point.}
A spin-glass practitioner will surely expect large statistical
fluctuations among different samples (recall that a sample is
characterized by a set of coupling constants $\{J_{ij}\}$). Probably,
large fluctuations would also be expected if different starting points
$\{\sigma_i(t=0)\}$ are compared for a given sample. What it is more
worrisome is that we did find large statistical fluctuations for a
given sample and starting point $\{J_{ij},\sigma_i(t=0)\}$, see
\figref{fig:fluctuaciones_q}. Unfortunately, this means that, in order
to characterize the conditional probability
\begin{equation}\label{eq:PJsigma0}
P[q(t)\,|\,\{J_{ij},\sigma_i(t=0)\}]
\end{equation}
one needs to consider a large number of trajectories, all of them
sharing the same starting configuration. Given our computational
resources, we had to restrict ourselves to 49 independent trajectories
per $\{J_{ij},\sigma_i(t=0)\}$. By analogy with the problem of
particle diffusion, one could think that the ensemble of these 49
trajectories encode the solution of the Fokker-Planck equation (rather
than the Langevin dynamics) for the conditional probability in
Equation~\eqref{eq:PJsigma0}. 

\begin{figure}[htbp]
\centering
\includegraphics[width=0.9\linewidth]{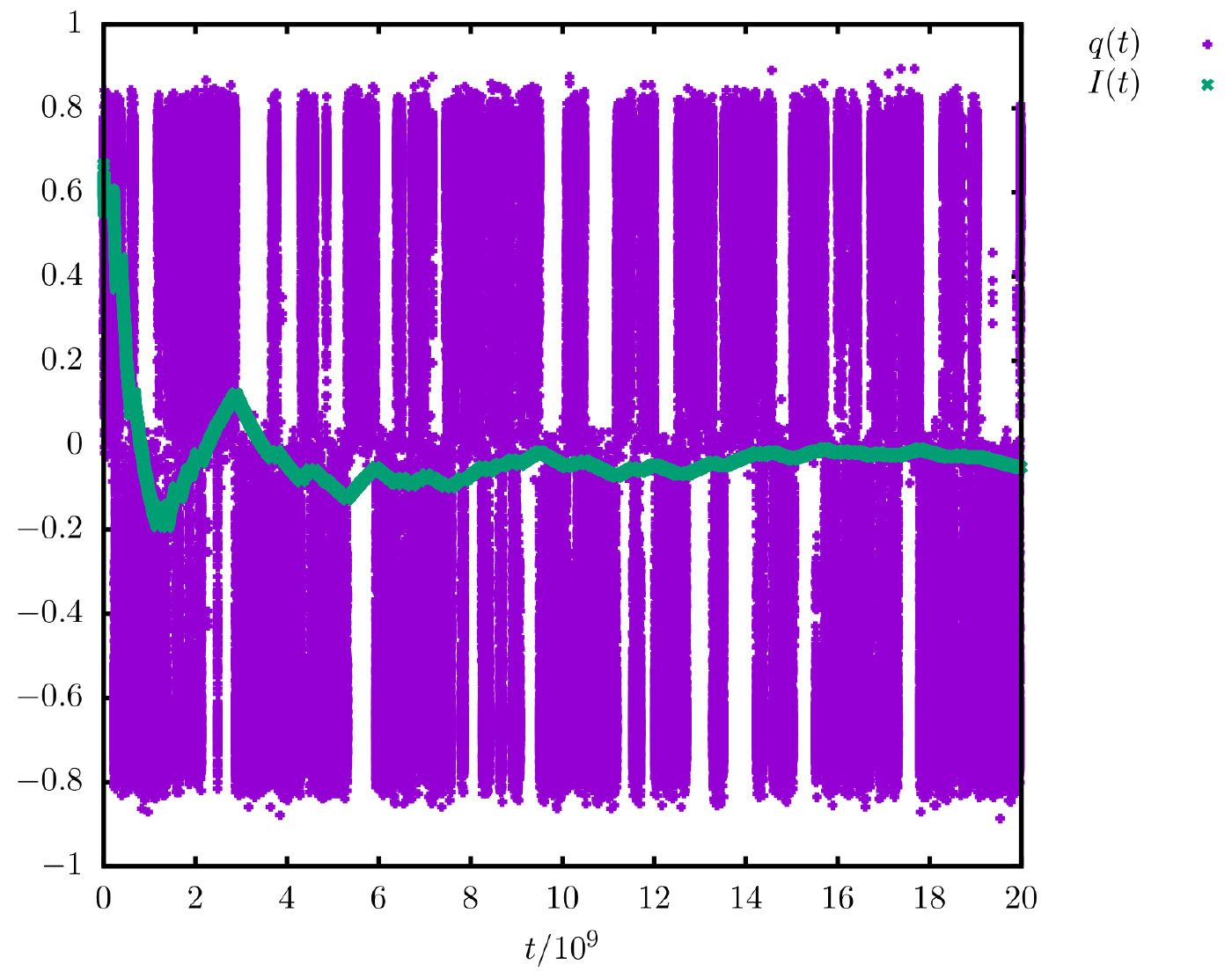}
\caption{Comparison of the time evolution $I(t)$ (green), recall
  Equation~\eqref{eq:I_t}, and $q(t)$ (purple), as computed for a
  randomly chosen sample for $\varepsilon=0$, from just one
  replica. We do not average over the 49 available replicas, in order
  to show that $I(t)$ displays a much softer time evolution.}
\label{fig:qt_vs_It}
\end{figure}

The large fluctuations of the conditional probability
$P[q(t)\,|\,\{J_{ij},\sigma_i(t=0)\}]$ is evinced in
\figref{fig:fluctuaciones_q} by the fact that the median and mean of
the distribution are wildly different. Furthermore, for this
particular sample, the median is ill-defined as the
$P[q(t)\,|\,\{J_{ij},\sigma_i(t=0)\}]$ is bimodal and gapped. We thus need
to further refine our dynamical study.

\paragraph{Smoothing $q(t)$ by time averaging.}
Clearly, we need to build from $q(t)$ a new observable, with less
fluctuations. To that end in mind, we considered the time average
\begin{equation}
I(t)\equiv\frac{\displaystyle\sum_{t'\leq t} \langle q_{t'}\rangle_r}{\displaystyle\sum_{t'\leq t} 1}\simeq\frac{1}{t}\int_0^t \langle q(t')\rangle_rdt'\,,
\label{eq:I_t}
\end{equation}
where now $\langle \ \rangle_r$ is the average over the 49 replicas, which reduces the fluctuations. 

In \figref{fig:qt_vs_It}, we compare $I(t)$ and $q(t)$, as computed
for a randomly chosen sample (in order to emphasize their different
behavior, $I(t)$ and $q(t)$ where computed from the very same \emph{single}
replica in \figref{fig:qt_vs_It}). We see that $I(t)$ shows a softer
time evolution. Furthermore, we see that short but inconsequential
excursions to distant states [which results in $q(t)<0$ for very short
  time intervals] have a smaller impact on $I(t)$.
\begin{figure}[htbp]
\centering
\includegraphics[width=0.9\linewidth]{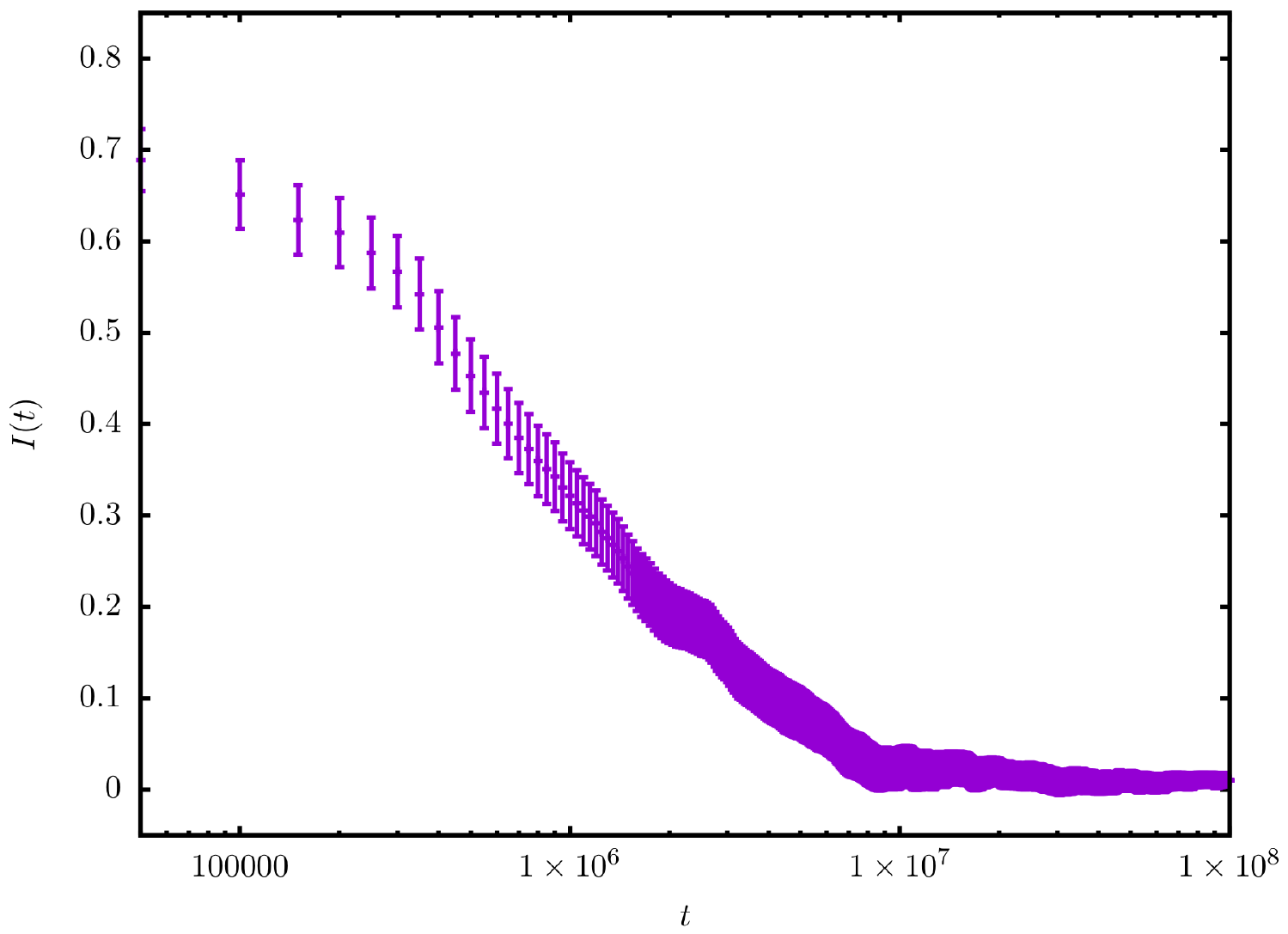}
\caption{The time evolution of the time-average $I(t)$, as computed
  for our sample \#50 at $\varepsilon=0$, is shown with its
  statistical error.  Even though (in the limit of an infinite number
  of replicas) $I(t)$ should be positive for all times, our
  49-replicas estimate becomes negative at finite $t$.  Therefore,
  estimating the relaxation time by requiring $I(t=\tau)=0$ would be a
  meaningless choice. In fact, $I(t=\tau)=0$ actually indicates the
  time when our signal-to-noise ratio becomes of order one.}
\label{fig:Elec_a}
\end{figure}
Now, we need to find a reasonable way to extract the relaxation time
$\tau$ from $I(t)$. One obvious solution is to require $I(t=\tau)=a$,
for some suitable value $a$. In order to chose $a$, we need some
reflection. On the one hand, $a$ should be significantly smaller than
$1$, to ensure that the system has escaped from the valley around its
starting configuration. On the other hand, a too small $a$ would
result in large errors. In fact, we expect for the replica-average of
$q(t)$\footnote{The characteristic times $\tau_n$ are obtained from
  the eigenvalues of the generator of the Markov-Chain
  $\lambda_n=\e^{-1/\tau_n}$~\cite{sokal:97}. For a system with $N$
  Ising spins there are $2^{N}$ characteristic times, the longest of
  which,  $\tau_1=\infty$ or $\lambda_1=1$, corresponds to the
  stationary measure. The reader will not that $\tau_1$ is absent from
  Eq.~\eqref{eq:E(q)} because $\langle q(t\to\infty)\rangle_r=0$.}
\begin{equation}
\langle q(t)\rangle_r = \sum_{n=2}^{2^N} a_n e^{-t/\tau_n}\,,
\label{eq:E(q)}
\end{equation}
which, when substituted in the integral form of \eqref{eq:I_t}, gives us
\begin{equation}
I(t)=\frac{1}{t}\sum_{n=2}^{2^N} a_n \tau_n\left[1-e^{-t/\tau_n}\right]\,.
\label{eq_E(I)}
\end{equation}
Now, because we expect that $a_n,\, \tau_n >0$, then
\begin{eqnarray}
I(t\rightarrow 0) & = & \sum_n a_n = 1 \,, \\
I(t\rightarrow \infty) & = & \frac{1}{t}\sum_n a_n\tau_n = 0\,,
\end{eqnarray}
in other words, $I(t)$ which does not vanish at any finite time. For
this reason, taking $a=0$, as \eqref{eq:cond_q(tau)} might suggest,
would implies that our statistical errors would equal (in order of
magnitude) the value of $I(t)$ itself. So we have to take a value of
$a$ far from $0$ which guarantees that the statistical errors do not
spoil our results, see \figref{fig:Elec_a}.

In order to make a reasonable choice for $a$, we consider a toy model
for $\langle q(t)\rangle_r$, with only two autocorrelation times
\begin{equation}
\langle q(t)\rangle_r^{\mathrm{toy}}=(1-q_{\mathrm{EA}})\,\mathrm{e}^{-t/\tau_{\mathrm{fast}}}\ +\ q_{\mathrm{EA}}\, \mathrm{e}^{-t/\tau_{\mathrm{slow}}}\,.
\end{equation}
If we now take the limit $\tau_{\mathrm{fast}}/\tau_{\mathrm{slow}}\to
0$, and recall $q_{\mathrm{EA}}\approx 0.844$, we immediately find
that requiring $I(t=\tau)=a$, with $a=0.437$ results in
$\tau=1.5\tau_{slow}$, which is a fairly sensible estimate of the
relaxation time. Furthermore, \figref{fig:Elec_a} tells us that
$a=0.437$ is sufficiently far away from zero to allow for a safe
determination of $\tau$, given the size of our statistical errors.

\section{\label{sec:ResulyDisc}Results and discussion}
\subsection{\label{sub:asignando_tiempos} On the computation of characteristic times for each sample}

\begin{figure}[htbp]
\begin{center}
\includegraphics[width=0.9\linewidth]{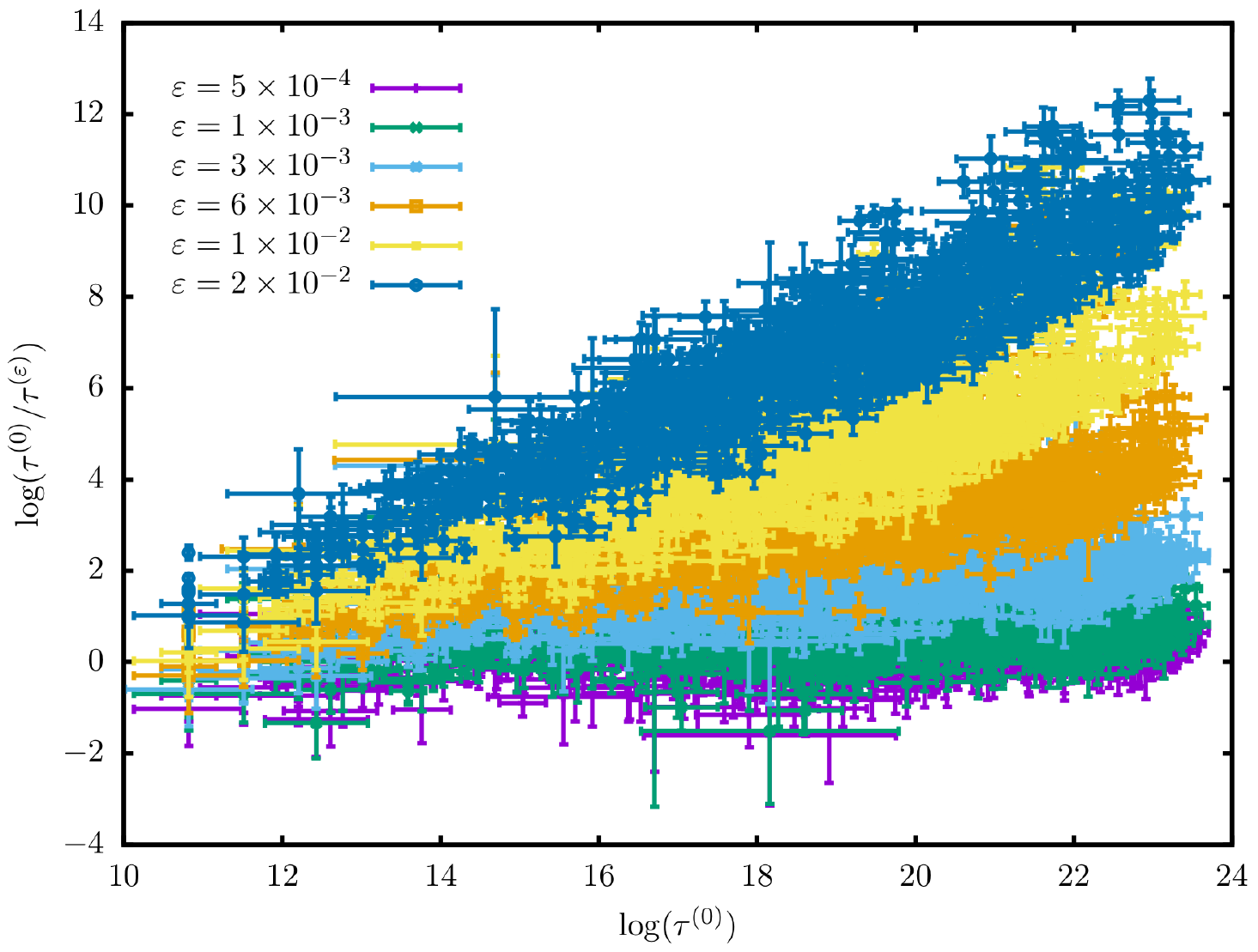}
\caption{$\log(\tau^{(0)}/\tau^{(\varepsilon)})$ vs $\log(\tau^{(0)})$
  for the different values of $\varepsilon$, as computed for all the
  pairs of samples and initial configurations that allowed us to
  estimate statistical errors through the Boostrap method explained in
  Section~\ref{sec:ResulyDisc} (we were able to compute errors for
  about 65\% of our 1280 samples).}
\label{fig:dif_vs_logt0}
\end{center}
\end{figure}

As we explained above, we compute a characteristic time scale $\tau^{(\varepsilon)}$ through the relation
\begin{equation}
I(\tau^{(\varepsilon)})=0.437\,.
\label{eq:condicion_tau}
\end{equation}
Note that every sample and starting configuration
$\{J_{ij},\sigma_i(t=0)\}$ has its own $\tau^{(\varepsilon)}$ for
every value of the external field $\varepsilon$. Now, it turns out
that the \emph{order of magnitude} of $\tau^{(\varepsilon)}$
fluctuates heavily among the different $\{J_{ij},\sigma_i(t=0)\}$ (see
\figref{fig:dif_vs_logt0}). Therefore, in order to properly quantify
the field-effect, we shall compute $\log
[\tau^{(0)}/\tau^{(\varepsilon)}]$ for every pair
$\{J_{ij},\sigma_i(t=0)\}$.

The main problems that we have needed to address upon computing $\log
[\tau^{(0)}/\tau^{(\varepsilon)}]$ have been the following:
\begin{enumerate}
\item For approximately a 35\% of pairs $\{J_{ij},\sigma_i(t=0)\}$,
  the condition \eqref{eq:condicion_tau} is not met for
  $\varepsilon=0$. This is hardly surprising, on the view of the
  sluggish dynamics illustrated in \figref{fig:DistribProb}. The same
  problem has occasionally arisen for $\varepsilon>0$, although to a
  lesser degree.
\item We need to estimate errorbars for both $\log\tau^{(0)}$ and $\log
[\tau^{(0)}/\tau^{(\varepsilon)}]$.
\end{enumerate}

To address the first problem, we have simply discarded from our
analysis all pairs $\{J_{ij},\sigma_i(t=0)\}$ for wich we could not
compute reliably either $\tau^{(0)}$ or
$\tau^{(\varepsilon)}$. Reliable here does not only mean that
condition \eqref{eq:condicion_tau} was met, but we also require that
errorbars could be computed (see below). Essentially, this amounts to
say that we are computing conditional 
probabilities over the pairs $\{J_{ij},\sigma_i(t=0)\}$, the imposed
condition being $\tau^{(0)}< 2\times 10^{10}$.

In order to compute errorbars for $\log\tau^{(0)}$ and $\log
[\tau^{(0)}/\tau^{(\varepsilon)}]$, we have employed a Bootstrap
method~\cite{efron:79}. Specifically, we have generated 1000 resampled
populations.  Each resampled population is obtained by choosing
randomly, with uniform probability, 49 trajectories out of our set of
49 trajectories (the same trajectory could be chosen several times, of
course). Then, by averaging over the 49 trajectories in the resampled
population, we recomputed $\langle q(t)\rangle_r$ and $I(t)$, and
obtained the corresponding $\log\tau^{(0)}$ and $\log
[\tau^{(0)}/\tau^{(\varepsilon)}]$. Now, we encountered one of three
alternatives:
\begin{itemize}
\item If we were able to compute $\log\tau^{(0)}$ and $\log
  [\tau^{(0)}/\tau^{(\varepsilon)}]$ for every one of the 1000
  resampled populations, then we assigned to that pair
  $\{J_{ij},\sigma_i(t=0)\}$ the values of $\log\tau^{(0)}$ and $\log
  [\tau^{(0)}/\tau^{(\varepsilon)}]$ obtained by averaging over the
  original set of 49 trajectories. Errors were computed by using
  standard Boostrap formulae.
\item If we were able to compute both $\log\tau^{(0)}$ and $\log
  [\tau^{(0)}/\tau^{(\varepsilon)}]$ for at least 84\% of the 1000
  resampled populations, then we assigned to that pair
  $\{J_{ij},\sigma_i(t=0)\}$ the median values of $\log\tau^{(0)}$ and
  $\log [\tau^{(0)}/\tau^{(\varepsilon)}]$. Errors were computed as
  the halved difference between percentiles 84 and 16.
\item We discarded pairs $\{J_{ij},\sigma_i(t=0)\}$ that did not
  fall in any of the two cases above.
\end{itemize}
The above analysis program produced values of $\log\tau^{(0)}$ and
$\log [\tau^{(0)}/\tau^{(\varepsilon)}]$, see
\figref{fig:dif_vs_logt0}, for 65\% of the pairs
$\{J_{ij},\sigma_i(t=0)\}$. The detailed analysis of the
$\varepsilon$-dependence will be addressed next.

\begin{figure}[t]
\begin{center}
\includegraphics[width=0.9\linewidth]{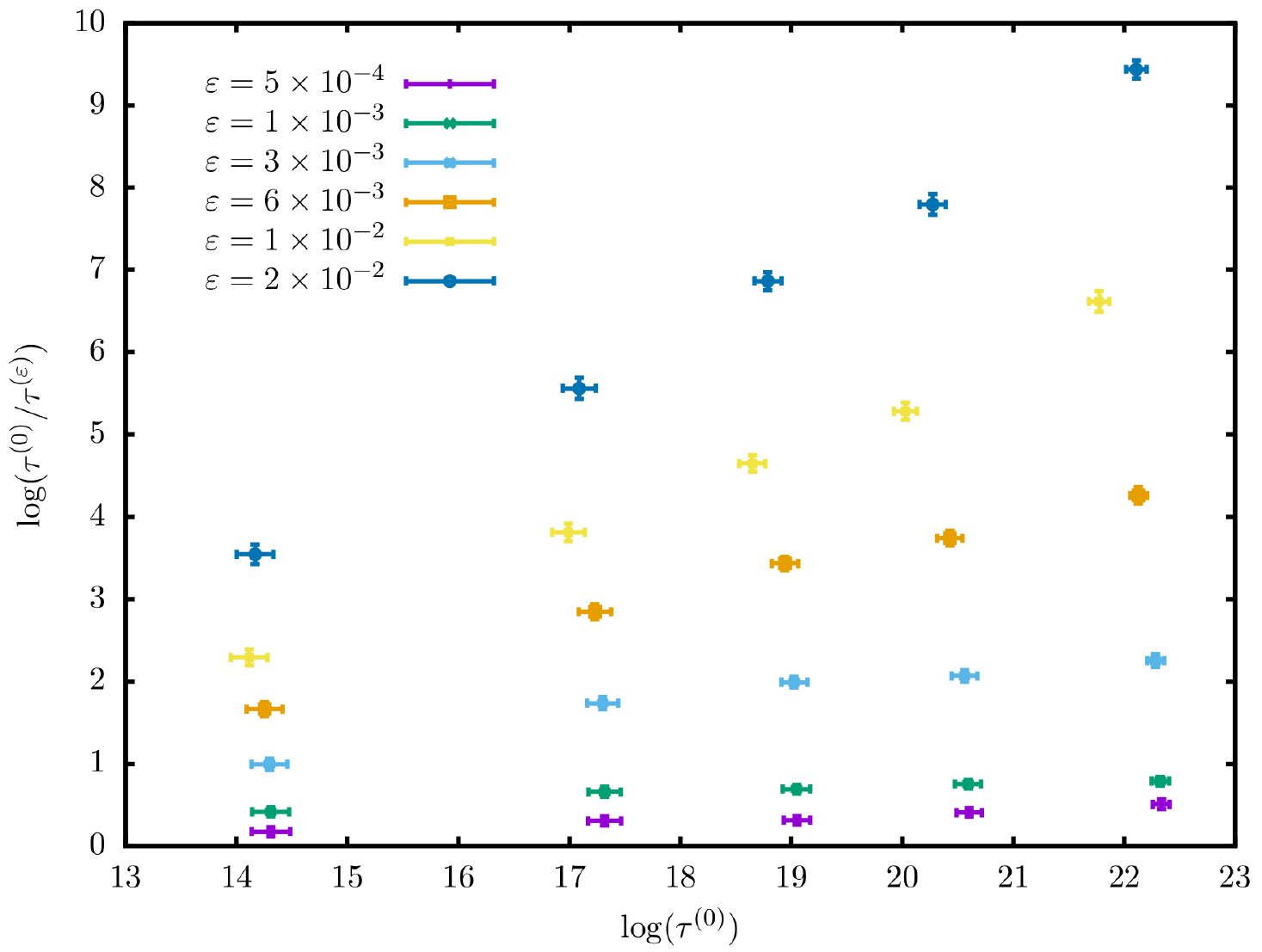}
\caption{The noise for
$\log(\tau^{(0)}/\tau^{(\varepsilon)})$ and $\log(\tau^{(0)})$ is considerably reduced, as compared to~\figref{fig:dif_vs_logt0}, by grouping the pairs of samples
and starting configurations in quintiles (see Section~\ref{sub:Resul} for details). The quintiles procedure does not arise the correlation between $\log(\tau^{(0)})$ and the response to the extenal field.}
\label{fig:quint_dif_vs_logt0}
\end{center}
\end{figure}
\subsection{\label{sub:Resul} On the relation between $\tau^{(\varepsilon)}$ and $\varepsilon$}

In spite of the strong fluctuations, we note an important feature in
\figref{fig:dif_vs_logt0}. Pairs $\{J_{ij},\sigma_i(t=0)\}$ with a
large $\log\tau^{(0)}$ are prone to a stronger reaction to
$\varepsilon$. In terms of the cartoon in
\figref{fig:E_con_sin_campo}, one would say that the larger the
barrier, the stronger the field-effect. This suggest already a
violation of the linear relationship between $\varepsilon$ and
$\log\tau^{(\varepsilon)}$ suggested in
Equation~\eqref{eq:tau_eps}. We shall further elaborate on
non-linearities in Section~\ref{sub:Virtual}.

In order to reduce the noise, while still respecting the strong
correlation between $\log\tau^{(0)}$ and the response to the field, we
have organized our pairs $\{J_{ij},\sigma_i(t=0)\}$ in quintiles
according to their $\log\tau^{(0)}$ value. Specifically, for every
value of $\varepsilon$, we have sorted our data according to
$\log\tau^{(0)}$. Next, we have grouped them in five sets of pairs
with the same number of elements. Obtaining exactly the same number of
elements in each group was impossible, because the total number of
pairs that met the criteria explained in
Section~\ref{sub:asignando_tiempos} was not a multiple of five, but
the differences among the cardinalities of the five sets was, at most, of
one element. The group containing the pairs $\{J_{ij},\sigma_i(t=0)\}$
with the smallest $\log\tau^{(0)}$ is referred to as quintile 1,
and so on. Our motivation for chosing quintiles rather than deciles (or even
a finer division), was keeping a large enough number of pairs within
in each group. The basic quantity that we have computed is the
arithmetic mean of both $\log\tau^{(0)}$ and $\log
[\tau^{(0)}/\tau^{(\varepsilon)}]$, for all pairs belonging to a
given quintile.

In order to estimates errors for the average $\log\tau^{(0)}$ and
$\log [\tau^{(0)}/\tau^{(\varepsilon)}]$ of each quintile, we used
again a boostrap method. We generated 1000 resampled populations by
choosing randomly, with uniform probability, 1280 pairs
$\{J_{ij},\sigma_i(t=0)\}$ among those in the original population
(note that the number of pairs that met the the criteria explained in
Section~\ref{sub:asignando_tiempos} varied among different ressampled
populations). In order to account as well for our errors in the
determination of $\log\tau^{(0)}$ and $\log
[\tau^{(0)}/\tau^{(\varepsilon)}]$ (these errors were due to our
limited number of trajectories for each pair), we modified the
estimates in the original population by adding two random numbers. For
each pair, these two random numbers were independent and normal
distributed, with zero average and dispersions equal to our errors for
$\log\tau^{(0)}$ and $\log [\tau^{(0)}/\tau^{(\varepsilon)}]$,
respectively. For every resampled population, we carried out the full
quintile procedure explained above. Errors for the average
$\log\tau^{(0)}$ and $\log [\tau^{(0)}/\tau^{(\varepsilon)}]$ of each
quintile were computed with the standard boostrap formula. The outcome of the quintile procedure is summarized in \figref{fig:quint_dif_vs_logt0}.

\begin{figure}[h]
\begin{center}
\includegraphics[width=0.9\linewidth]{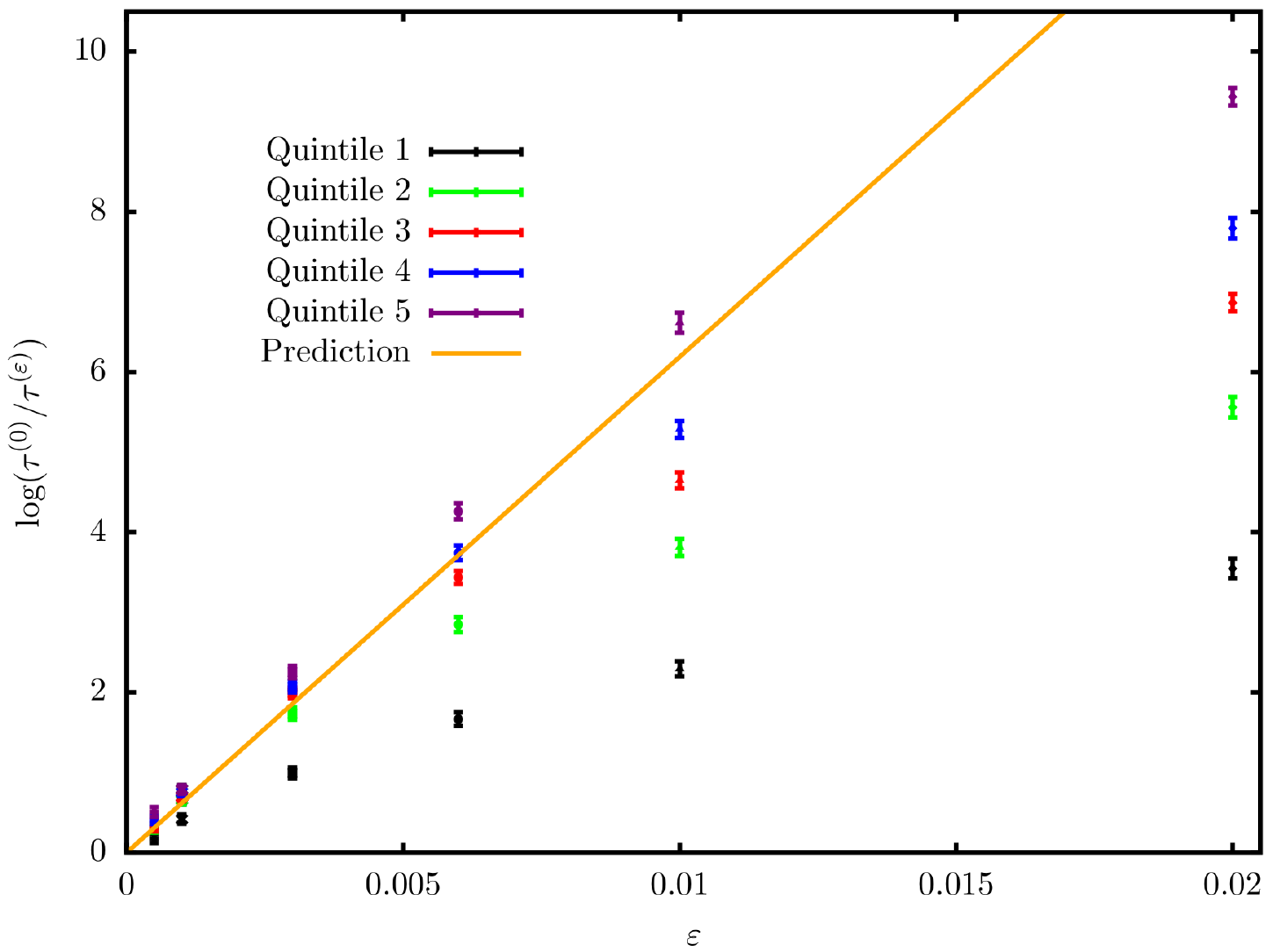}
\caption{Quintile averaged $\log(\tau^{(0)}/\tau^{(\varepsilon)})$ as
  a function of the extenally applied field $\varepsilon$ (data were
  taken from \figref{fig:quint_dif_vs_logt0}). The orange line is
  given by Equation \eqref{eq:tau_eps}, where we take $\Delta q=
  q_{\mathrm{EA}}\approx 0.844$. Recall that $q_{\mathrm{EA}}$ is the
  position of the maximum of the probability density shown in
  \figref{fig:DistribProb}.}
\label{fig:dif_vs_eps}
\end{center}
\end{figure}
Next, we consider the quintile-average of
$\log(\tau^{(0)}/\tau^{(\varepsilon)})$ as a function the externally
applied field, see \figref{fig:dif_vs_eps}. We note the following features:
\begin{itemize}
\item For all five quintiles, and for small-enough fields, we can
  identify in \figref{fig:dif_vs_eps} a linear relationship between
  $\varepsilon$ and $\log\tau^{(\varepsilon)}$. Furthermore, the
  slopes close to $\varepsilon=0$ approach the predicted one (orange
  line in \figref{fig:dif_vs_eps}). Nevertheless, the region of
  validity of the linear approximation is strongly
  quintile-dependent. In particular, for the slowest pairs
  $\{J_{ij},\sigma_i(t=0)\}$, the region where the linear
  approximation is reasonable extends up to $\varepsilon\approx 0.01$.
  This implies that we could have guessed the behavior for smaller
  $\varepsilon$ with far lesser numerical effort.
\item In addition, the slowest pairs $\{J_{ij},\sigma_i(t=0)\}$ show a
  reduction of three orders of magnitude in their relaxation time for
  $\varepsilon$ near $0.01$. Notice that, at $\varepsilon=0$, the
  fastest pairs are thousand times faster than the slowest
  pairs. Therefore, the external field produces an homogenization of
  the relaxation times (homogenization in order of magnitude, at least).
\item However, we note that for some quintiles the external field
  increases $\log(\tau^{(0)}/\tau^{(\varepsilon)})$ \emph{above} the
  prediction by the linear approximation, see the orange line in
  \figref{fig:dif_vs_eps}. Clearly, some unsuspected processes enlarge the ability of the field to reduce the relaxation times. We speculate in Section~\ref{sub:Virtual} about one such escaping mechanism.
\end{itemize}

\subsection{\label{sub:Virtual}Virtual states}

We discuss here how virtual states might produce an enhanced scape
rate from the starting configuration, in the presence of an external
field. In order to do that, we first need to relate the calculation
process that we use to estimate $\tau$ with the effective potential.

\begin{figure}
\centering
\begin{tikzpicture}[domain=-2.05:2.05,range=-1.1:1.1,smooth,scale=1.3]
\draw (0,-1.1) -- (0,1.1);
\node at (0,1.1) [anchor=south east] {$\Omega'(q)$};
\draw[color=blue,line width=1pt] (-2,-0.6)--(2,-0.6);
\draw (-2.2,0) -- (2.2,0);
\node at (2.2,0) [anchor= west] {$q$};
\node at (2.2,-0.6) [anchor=north west,color=blue] {$\varepsilon$};
\node at (-1.95,-2pt) [anchor=south east] {$-1$};
\node at (1.95,-2.2pt) [anchor=south west] {$1$};
\draw[clip] plot (\x,{2*(-2.8*\x/2+14*(\x/2)^3-25*(\x/2)^5+14*(\x/2)^7)});
\draw (-2,-2pt) -- (-2,2pt);
\draw (2,-1pt) -- (2,1pt);
\draw (1.25,0.5) -- (1.25,0);
\fill[black] (1.25,0) -- (1.3,0.1) -- (1.2,0.1) -- cycle;
\fill[green!20] (0,0) rectangle (2,-0.6);

\fill[red!20] (0,-0.6) rectangle (2,-2);
\draw[color=blue,line width=1pt] (-2,-0.6)--(2,-0.6);
\end{tikzpicture}
\caption{Schematic representation of the derivative of the
  Franz-Parisi-like potential in Equation~\eqref{eq:P(q)}. Stationary
  points are given by the condition $\Omega'(q)=-\varepsilon$ (recall
  that $\varepsilon$ is a repulsion). Local minima verify in
  addition that $\Omega''(q)>0$. At $\varepsilon=0$ we encounter two
  local minima separated by a saddle point at $q=0$. Instead, at
  $\varepsilon>0$ we find \emph{three} local minima separated by two
  saddle points. The arrow indicates a characteristic value of $q$, employed in the computation of $\tau$ (see text).}
\label{fig:est_virtuales}
\end{figure}
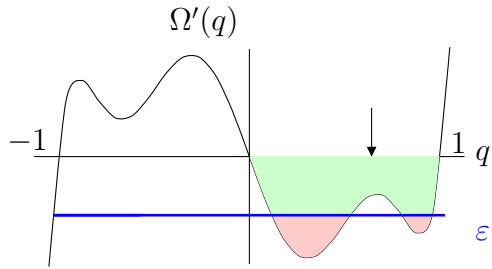

Consider a sample and a starting configuration at $\varepsilon=0$. Let
the time elapsed from system prepation go to infinity, so that all
statistical correlations between the initial and the current
configuration fade away. Under such circumstances, the probability
density $P(q)$ for the overlap $q$ between the starting and the
current configuration defines the effective potential $\Omega(q)$
through
\begin{equation}
P(q) \propto e^{-N\Omega(q)}\,,
\label{eq:P(q)}
\end{equation}
(the normalization is not relevant now). Keep in mind that $\Omega(q)$
depends not only on the sample, but on the starting configuration as
well (in fact, $\Omega(q)$ can be regarded as the Franz-Parisi
potential~\cite{franz:97}). For the purpose of discussion, let us
consider an effective-potential profile like the one in
\figref{fig:est_virtuales}, where we plot the derivative of the
effective potential as a function of the correlation. If we maximize
the probability given by equation \eqref{eq:P(q)}, we find that the
equilibrium states satisfy $\Omega'(q)=0$ and $\Omega''(q)>0$, so we
have two symmetric real states at $q\approx \pm q_{\mathrm{EA}}$,
separated by a saddle-point at $q=0$.

Now, our procedure for computing $\tau$ declares that the system has
\emph{escaped} from the initial valley at $q\approx q_{\mathrm{EA}}$
when the overlap $q(t)$ goes below some threshold, indicated by an
arrow in \figref{fig:est_virtuales}. In fact, we know that the system
will be tethered to $q\approx q_{\mathrm{EA}}$ until a random
excursion will take it to the saddle-point at $q=0$. Once the
saddle-point is reached, the transit to the minimum at
$q\approx -q_\mathrm{EA}$ is very fast. Therefore, the logarithm of
$\tau^{(\varepsilon=0)}$ is given by the barrier ($B$), namely by the
area between the $q$ axis and the function $\Omega'(q)$ (filled area
in \figref{fig:est_virtuales}).

In the presence of an external field, the local minima are given by
the conditions $\Omega'(q)=-\varepsilon$ and
$\Omega''(q)>0$. Therefore, the two original minima at $q\approx \pm
q_{\mathrm{EA}}$ get small corrections of order $\varepsilon$. It is
more significant, however, that new states can appear for
$\varepsilon>0$, such as the one depicted in the figure. We name this
new minimum a \emph{virtual state}. Now, if the arrow in
\figref{fig:est_virtuales} lies to the right of the virtual state, the
most probable process for escaping the original minimum at $q\approx
q_{\mathrm{EA}}$ is not jumping to the symmetric state at $q\approx
-q_{\mathrm{EA}}$, but stopping at the virtual state. The
corresponding barrier is given by the area of the \emph{right} pink
region in \figref{fig:est_virtuales}.  Instead, the relevant barrier
for the escape process to $q\approx -q_{\mathrm{EA}}$ (which is the
only escape path at $\varepsilon=0$) is given by the area of the
\emph{two} pink regions in the figure. In other words,
$\log\tau^{(\varepsilon)}$ will be quite smaller than what one would
have guessed from $\log\tau^{(\varepsilon=0)}$.

\begin{figure}[htbp]
\centering
\includegraphics[width=0.9\linewidth]{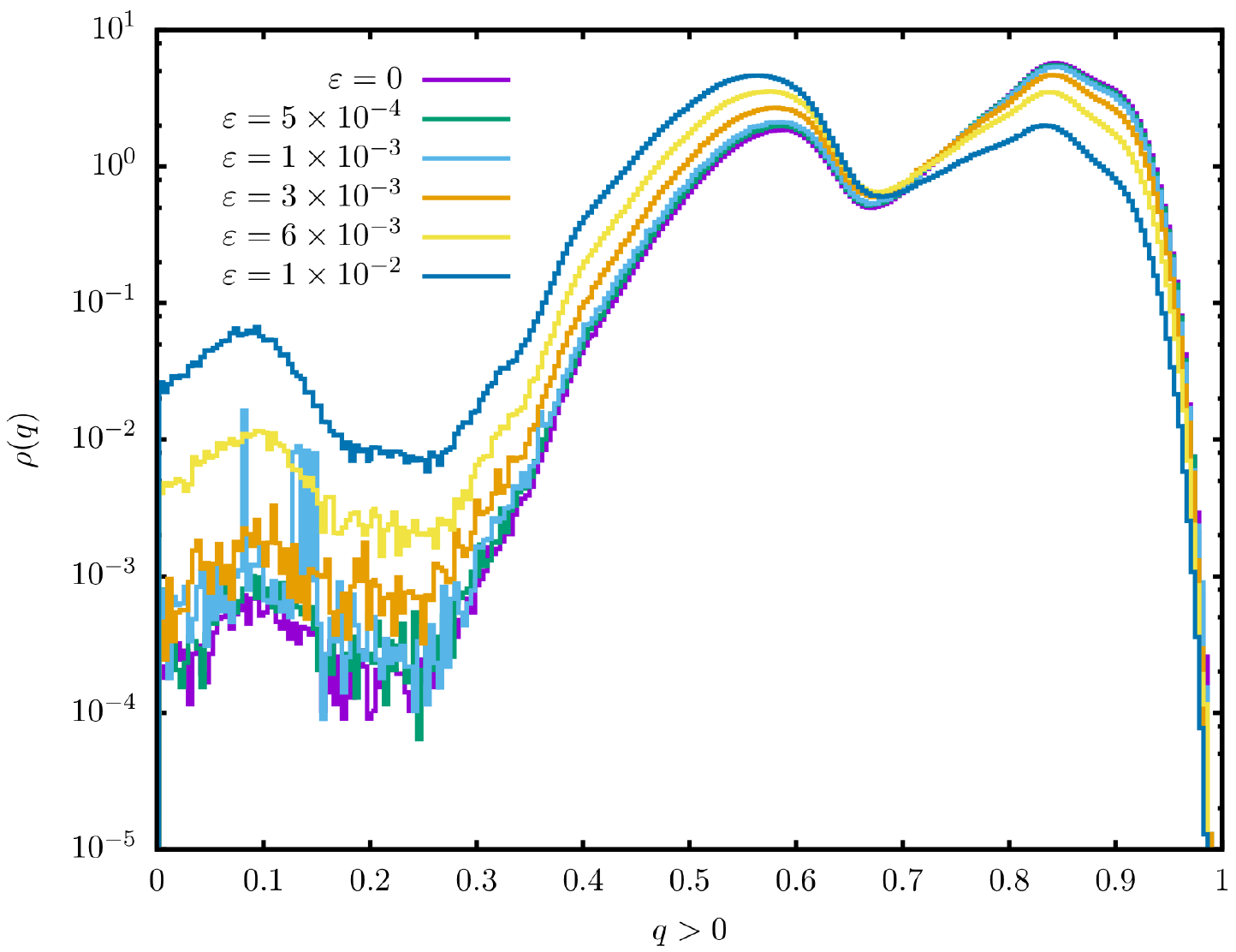}
\caption{Semilogarithmic plot of the probability density conditional to
$q(t)>0$, for a sample and starting configuration remarkable for its large value of $\log \tau^{(\varepsilon=5\times 10^{-4})}/\log\tau^{(0)}$. We first computed
the conditional probability density for each replica, and only afterwards averaged over the 49 trajectories. Besides the local maxima at $q\approx 0.84$ and $q\approx 0.6$, which are almost unaffected by the external field, a new state
 (a virtual state, probably) emerges at $q=0.1$ upon increasing $\varepsilon$.}
\label{fig:prob_virtual}
\end{figure}

In order to give some flesh to the virtual-state idea, we show
in~\figref{fig:prob_virtual} the probability density of $q(t)$,
conditional to $q(t)>0$ for one particular sample and starting
configuration, chosen because of its anomalously large ratio $\log \tau^{(\varepsilon=5\times 10^{-4})}/\log\tau^{(0)}$. At $\varepsilon=0$ we observe two
well defined states (i.e. local maxima of the probability density)
at $q\approx 0.84$ and $q\approx 0.6$. When $\varepsilon$ increases,
a new local maximum (a virtual state?) emerges at $q\approx 0.1$.

\section{\label{sec:Concl}Conclusion}
The huge relaxation times of glasses make challenging their
study. This difficulty has prompted researchers to invent
non-physical dynamics, correctly sampling the Boltzmann weight even at
low temperatures (see
e.g. Refs.~\cite{hukushima:96,marinari:98b,grigera:01,fernandez:06,ninarello:17}). However,
reaching equilibrium is only half of the problem when (as it is the
case for many glass formers), the glass transition is not accompanied
by significant changes in static quantities, think for instance of
structure factors or density profiles. Under such circumstances, one
is forced to study Nature-imitating dynamics starting from equilibrium
configurations. Dynamics of this kind are extremely slow.

A method to reduce free-energy barriers, hence accelerating relaxations,
consists in placing the system in an external field.  This idea has
been used experimentally to extract the spin-glass correlation
length~\cite{joh:99,guchhait:17} by placing the spin-glass in an
external magnetic field. Similar studies are being conducted for the
dielectric response of glass-forming
liquids~\cite{lhote:14,ladieu:12}. Here, we perform a preliminary
exploration of this strategy in the highly controlled context of the
simulation of a small spin-glass system in three dimensions. Let us recall
that our external field $\varepsilon$ should be proportional to
the \emph{square} of the external fields used in experiments.

Our simulations have shown that the external field indeed reduces the
relaxation times, which can be regarded as a reduction of the height
of the effective barriers (recall the cartoon in
\figref{fig:E_con_sin_campo}). Nevertheless, we have found rather
dramatic statistical fluctuations. We have needed to resort to robust
statistical methods (i.e. studying quintiles), in order to control
these fluctuations. After taking care of fluctuations, we have found
that our naive expectation of a linear dependence on the external
field of the logarithm of the relaxation time is only full-filed for
very small fields. 

When regarded as a purely numerical strategy, we have found that the
external field may reduce the relaxation times by some three orders of
magnitude. Furthermore, the field strongly enhances the homogeneity of
the relaxation times of the different samples and starting
configurations.

Finally, we have discussed a possible mechanism (namely the formation
of a virtual state), to explain the extreme sensitivity to the
external field of some samples and starting configurations.

\ack We thank E. Marinari, D. Yllanes and F. Ladieu for encouraging
discussions.  This work was partially supported by Ministerio de
Econom\'ia, Industria y Competitividad (MINECO) Grants
No. FIS2015-65078-C2, No. FIS2016-76359-P (both partly funded by
FEDER). We were also partly funded by the Junta de Extremadura (Spain)
through Grants No.  GRU10158 and IB1603 (partially funded by FEDER).
This project has received funding from the European Research Council
(ERC) under the European Union's Horizon 2020 research and innovation
program (Grant No. 694925 GlassUniversality).

\appendix

\section{\label{Ap:Sim} Multispin coding at fixed temperatures}

Our multisample multispin coding Parallel Tempering simulation is
fully standard and does not deserve any special comment. However, we
have found it useful to describe here our simulation methods at fixed
temperature in the presence of the external field, where random
numbers are \emph{not} recycled in the simulation of the different
bits in a computer word.

As usual in multispin coding simulations, the key of our program is
that we can use a binary representation of $J_{ij}$ and $\sigma_i$,
namely $J_{ij}=\left\lbrace-1,1\right\rbrace\rightarrow
b^J_{ij}=\left\lbrace 1,0\right\rbrace$ and
$\sigma_i=\left\lbrace-1,1\right\rbrace\rightarrow
b^{\sigma}_i=\left\lbrace 0,1\right\rbrace$. Multiplications of spin
variables, for instance, are equivalent to the XOR boolean
operation. Furthermore, we exploit that boolean operations (such as
AND, OR, XOR, NOT,\ldots) are carried out in parallel for all bits in
a computer word.

We employ a modified version of the Metropolis algorithm for a
spin-system in a field, which is particularly well suited for
multi-spin coding simulations. Let us attempt to flip a spin, the
corresponding energy change is composed of two terms, $\Delta E =
\Delta E_J + \Delta E_\varepsilon$. The first term $\Delta E_J\in
\{0,\pm 4, \pm 8, \pm 12\}$ is the energy change due to the exchange
term in the Hamiltonian in Equation~\eqref{eq:repulsion}. On the other
hand, $ \Delta E_\varepsilon= -2\varepsilon\sigma_i(t)\sigma_i(t=0)$
is the energy change due to the external field. Hence, the spin flip
is accepted with a probability which is a product of two terms\footnote{The most often used version of the Metropolis probability is $\min\left\lbrace 1, \exp(-[\Delta E_J+\Delta E_\varepsilon]/T)\right\rbrace$, but Eq.~\eqref{eq:nuestro-Metropolis} is physically equivalent and more convenient for us.}
\begin{equation}\label{eq:nuestro-Metropolis}
\mathrm{Prob[spin\ flip]} =
\min\left\lbrace 1, \exp(-\Delta E_J/T)\right\rbrace\, \cdot\min\left\lbrace 1, \exp(-\Delta E_\varepsilon/T)\right\rbrace\,.
\end{equation}
In practice, the Monte Carlo dynamics is implemented as
\begin{eqnarray}
b^{\sigma}_i(t+1)&=&b_i^{(\mathrm{change})}\,\mathrm{XOR}\, b^{\sigma}_i(t)\,,\\[1mm]
b_i^{(\mathrm{change})}&=&b_i^{(\mathrm{change},J)}\,\mathrm{AND}\,b_i^{(\mathrm{change},\varepsilon)}\,,
\end{eqnarray}
with
\begin{eqnarray}
\mathrm{Probability}\Big[b_i^{(\mathrm{change},J)}=1\Big]&=& \min\left\lbrace 1, \exp(-\Delta
E_J/T)\right\rbrace\,,\\
\mathrm{Probability}\Big[b_i^{(\mathrm{change},\varepsilon)}=1\Big]&=& \min\left\lbrace 1, \exp(-\Delta
E_\varepsilon/T)\right\rbrace\,.
\label{eq:b_cambio}
\end{eqnarray}
In order to avoid unwanted correlations, the 256 bits used to update a
computer word, namely 128 bits $b_i^{(\mathrm{change},J)}$ and 128
bits $b_i^{(\mathrm{change},\varepsilon)}$, should be statistically
independent. A naive implementation of the method would require
getting 256 independent random numbers per computer word, which is
rather costly. Fortunately, there is a better way.

For the obtention of the bits $b_i^{(\mathrm{change},J)}$, we employ
the Daemons algorithm by Ito and Kanada~\cite{ito:90} which, up to our
knowledge, needs a smaller number of Boolean operations than any other
method. As for the random numbers, we observe that the smallest
exchange-energy barrier to be overcome, namely $\Delta E_J=4$, is rather large
as compared to our working temperature $T=0.698$. Hence, only very
rarely the thermal bath allows us to overcome any energy barrier, and
a Gillespie-Bortz method is called for~\cite{bortz:75,gillespie:77}. 

For the sake of completeness, let us recall how we use the
Gillespie-Bortz method to run our Monte Carlo algorithm for long times
without the need to generate a large number of random numbers. Indeed,
in a naif simulation, one would draw for each bit an independent,
uniformly distributed random number $0<R<1$. The corresponding bit is
set to 1 only if $R< p$, where $p$ is some probability. The crucial
observation is that, if $p$ is small enough (in our case $p\leq
\mathrm{e}^{-4/T}=0.0032\ldots$), almost every bit will be zero with
very few exceptions. The Gillespie-Bortz method allows to correctly
select those exceptions, with very little effort.  Let us imagine that
we have just set the bit to one, the probability that the next bit set
to one will be found after $k$ independent extractions of the random
numbers $R$ is $(1-p)^{k-1}p$.  What we do is to extract $k$ according
to that probability. Indeed, the typical value of $k$ is $1/p$. So, in
our case, we can obtain approximately 308 random bits from a single extraction
of $k$. For more details about the implementation of these ideas, the
reader may wish to check Ref.~\cite{fernandez:15}.

As for the bits $b_i^{(\mathrm{change},\varepsilon)}$, the
straigtforward implementation is
\begin{equation}
b_i^{(\mathrm{change},\varepsilon)}=\Big[\,\mathrm{NOT}[\,b_i^{(0)}\,\mathrm{XOR}\, b_i^{(t)}\,]\,\Big]
\ \mathrm{OR}\  [R<\exp(-2\varepsilon/T)] \,,
\end{equation}
where $0<R<1$ is an uniformly distributed random number, and the
corresponding bit is set to one if the inequality is
fulfilled. However, the probability $\exp(-2\varepsilon/T)$ is close
to 1, due to the smallness of $\varepsilon$, which  would preclude us from using Gillespie methods. Fortunately, the problem is fairly easy to fix:
\begin{eqnarray}
b_i^{(\mathrm{change},\varepsilon)}=&&\\
\Big[\,\mathrm{NOT}[\,b_i^{(0)}\,\mathrm{XOR}\, b_i^{(t)}\,]\,\Big]
\ \mathrm{OR}\ \ \Big[\,\mathrm{NOT}\,[\,R < (1-\exp(-2\varepsilon/T))\,] \,\Big]&&\,.\nonumber
\end{eqnarray}

\section*{References}

\bibliographystyle{unsrt.bst}
\bibliography{/homenfs/rg/biblio}

\end{document}